\documentclass[12pt]{article}
 \pdfoutput=1
 \usepackage{graphicx}
  \usepackage{hyperref}
  \usepackage{epsfig}
  \usepackage{amsmath}
  \usepackage{amssymb}
  \usepackage{color}
  \usepackage[nosort]{cite}
  \newlength{\abstractwidth}
  \newcommand{\be}{\begin{equation}}
  \newcommand{\ee}{\end{equation}}
  \newcommand{\tr}{\text{tr}}
  \renewcommand{\title}[1]{\vbox{\center\bf{\Large{#1}}}\vspace{5mm}}
  \renewcommand{\author}[1]{\vbox{\center#1}\vspace{5mm}}
  \newcommand{\address}[1]{\vbox{\center\em#1}}
  \newcommand{\email}[1]{\vbox{\center\tt#1}\vspace{5mm}}

 \newcommand{\bea}{\begin{eqnarray}}
  \newcommand{\eea}{\end{eqnarray}}
  \def\nref#1{(\ref{#1})}
  \def\la{\label}

\begin{document}

\begin{titlepage}
\begin{center}
\hfill \\
\hfill \\
\vskip 1cm

\title{A bound on chaos}

\author{Juan Maldacena${}^1$, Stephen H. Shenker${}^2$ and Douglas Stanford${}^{1}$}

\address{
 ${}^1$ School of Natural Sciences, Institute for Advanced Study \\
 Princeton, NJ, USA\\
  ${}^2$ Stanford Institute for Theoretical Physics {\it and} \\
 Department of Physics, Stanford University \\
 Stanford, CA, USA

}

\email{malda@ias.edu, sshenker@stanford.edu,
stanford@ias.edu}

\end{center}
  
  \begin{abstract}
  
We conjecture a sharp bound on the rate of growth of chaos in thermal quantum systems with a large number of degrees of freedom.  Chaos can be diagnosed using an out-of-time-order correlation function closely related to the commutator of operators separated in time. We conjecture that the influence of chaos on this correlator can develop no faster than exponentially, with Lyapunov exponent $\lambda_L \le 2 \pi k_B T/\hbar$. We give a precise mathematical argument,  based on plausible physical assumptions, establishing this conjecture.
  \end{abstract}
   \end{titlepage}

\tableofcontents

\baselineskip=17.63pt

\section{Introduction}

Strong chaos, the butterfly effect,  is a ubiquitous phenomenon in physical systems, explaining thermal behavior, among other things.  In quantum mechanics, this phenomenon can be characterized using the commutator $[W(t),V(0)]$ between rather general Hermitian operators at time separation $t$. The commutator diagnoses the effect of perturbations by $V$ on later measurements of $W$ and vice versa. One indication of the strength of such effects is
\be
C(t) = -\langle[W(t),V(0)]^2\rangle\label{C}
\ee
where $\langle \cdot\rangle = Z^{-1}\tr[e^{-\beta H}\cdot]$ denotes the thermal expectation value at temperature $T = \beta^{-1}$. A quantum definition of the butterfly effect is that $C(t)$ should become of order $2 \langle VV\rangle\langle WW\rangle$ for large $t$, regardless of the specific choice of $V,W$ \cite{Almheiri:2013hfa} within an appropriate class. In general, we assume $V(0)$ and $W(0)$ are simple Hermitian operators, describable as a sum of terms, each a product of only ${\cal O}(1)$ degrees of freedom.\footnote{Traces of finite products of matrix fields are simple by this definition; time evolved operators $e^{-iHt}Oe^{iHt}$ with $t$ large are generally not.} We further assume that $V,W$ have zero thermal one point functions.

We call the time scale where $C(t)$ becomes significant the ``scrambling time" $t_*$  \cite{Hayden:2007cs,Sekino:2008he}. There is another shorter time scale relevant for chaos, the exponential  decay  time $t_d$ for two point expectation values like $\langle V(0) V(t) \rangle$.  We call this time scale the ``dissipation time," or, when a quasiparticle description applies, the ``collision time."  In the strongly coupled systems we will focus on, we expect $t_d \sim \beta$. We also expect general time ordered correlators to approach their long time limits after this time scale.  For example 
$\langle V(0)V(0) W(t) W(t) \rangle \sim \langle V V \rangle \langle W W \rangle + {\cal O}(e^{-t/t_d})$. 

We can gain some intuition for the relation between $C(t)$ and chaos by studying the semiclassical limit of a one particle quantum chaotic system, like semiclassical billiards, following the classic reference \cite{larkin}.  Schematically, in the semiclassical limit taking $V = p$ and $W(t) = q(t)$ the commutator $[q(t), p]$ becomes the Poisson bracket $i \hbar \{q(t), p\} = i \hbar \frac{\partial q(t)}{ \partial q(0)}$.  This gives the dependence of the final position on small changes in the initial position, the classical diagnostic of the butterfly effect.   Nearby trajectories in such systems diverge exponentially, $\sim e^{\lambda_L t}$ where $\lambda_L$ is a Lyapunov exponent.    Here $t_d \sim {1 \over \lambda_L}$.   For early times, the correlator $C(t) \sim \hbar^2 e^{2 \lambda_L t}$ so $t_* \sim {1 \over \lambda_L} \log {1 \over \hbar}$.  In this context $t_*$ is called the ``Ehrenfest time."  There is a parametrically large hierarchy between scrambling and collision times determined, in this case, by the small parameter $\epsilon = \hbar$.   Systems with such a large hierarchy will be the focus of this paper.

From a purely quantum mechanical point of view we can follow the analysis of \cite{RSS} and use $C(t)$ as a measure of the  growth of the operator $W(t)$  expressed as a sum of products of simple basis operators.  In qubit models these would just be Pauli matrices.  A large commutator indicates a complicated operator $W(t)$  that arises because chaos disrupts the cancellation between the initial and final factors in $W(t) = e^{i H t} W e^{-iHt}$.  If the number of qubits $N_q$ is large it will in general take a long time for a large commutator to build up.  If the interactions are local, the time is linear in the separation between $W$ and $V$ \cite{Lieb:1972wy,Hastings:2005pr,hastingslocality}. Even if the interactions are nonlocal, but are formed from products of just a few qubits,  it will take a time $t_* \sim \log N_q$ for $C(t)$ to become large.\footnote{Scrambling in nonlocal quantum circuits was studied in \cite{dankert:2009a,harrow:2009a,arnaud:2008a,brown:2010a,diniz:2011a,Brown:2012gy}. A logarithmic scrambling time was conjectured for nonlocal Hamiltonian systems in \cite{Hayden:2007cs, Sekino:2008he}, and supported by a Lieb-Robinson bound in \cite{Lashkari:2011yi}.} The analog of $t_d$ here is roughly the time for $W(t)$ to add a few Pauli matrices, so large $N_q$ qubit systems provide another example of a large hierarchy between $t_*$ and $t_d$. Clearly these ideas  generalize to a wide variety of lattice quantum systems.  Here $1/\epsilon$ would be the size of the system (in the nonlocal case), or an exponential of the distance between the $V(0)$ and $W(0)$ operators (in the local case).

\begin{figure}
\begin{center}
\includegraphics[scale = 0.7]{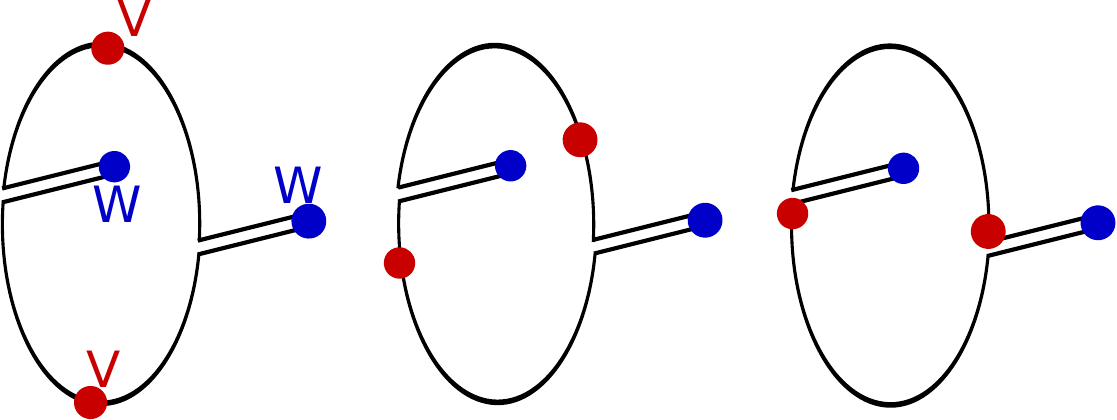}
\caption{$F(t)$ is the correlation function of operators arranged on the thermal circle as shown at left. The folds indicate the Lorentzian time evolution to produce $W(t)$. At complex time, $F(t+i\tau)$ is given by rotating one of the pairs of operators by angle $2\pi\tau/\beta$. The center panel corresponds to $|\tau| < \beta/4$ and the right panel corresponds to $\tau = \beta/4$.}\label{fig1}
\end{center}
\end{figure}

In a lattice system, the square of the commutator in $C(t)$ is a reasonable operator, but in a quantum field theory it generally requires regularization. A convenient prescription is to move one of the commutators halfway around the thermal circle, so that we consider
\be
-\tr\left[y^2 [W(t),V]y^2 [W(t),V] \right]
\ee
where $y$ is defined by
\be
y^4 = \frac{1}{Z} e^{-\beta H},
\ee
and $V$ is always $V(0)$. A closely related function, and the one that we will work with directly in this paper, is
\be
F(t) = \tr[ yVyW(t)yVyW(t)],\label{F}
\ee
corresponding to insertion of the $V$ and $W$ operators at equal spacing around the thermal circle. As explained in Fig.~\ref{fig1}, $F$ is analytic in a strip of width $\beta/2$ in the complex time plane, and at the edges of this strip we can relate $F$ to the regularized commutator discussed above. To see this, notice that $F(t - i\beta/4) = \tr[ y^2 V W(t) y^2 V W(t)]$, so
\bea
- \tr\left[y^2 [W(t),V]y^2 [W(t),V] \right] &=& \tr[ y^2 W(t) V y^2 V W(t)] + \tr[ y^2 V W(t)  y^2  W(t) V]  
\cr
&&~~  -F(t + i{ \beta \over 4}) - F( t - i {\beta \over 4}). ~~~~
\eea

We can use this equation to develop some intuition for the time dependence of $F(t)$. First, for small $t$, all terms on the RHS are positive and roughly equal. The commutator is small because of a cancellation between the first and second lines. The terms on the first line can be  be written as norms of states, e.g. $ y W(t) V y^{-1} |TFD\rangle$ (the state $|TFD\rangle$ is defined below), so they  remain of order one at large $t$. The growth of the commutator is therefore due to a decrease in $F(t \pm i \beta/4)$. This gives us a second quantum definition of the butterfly effect: at large $t$, $F$ should become small, regardless of $V,W$.

We will give two additional pieces of intuition for the late-time decrease of $F$. The first is based on the observation that, as $t$ becomes large, all pairs of operators are separated by large intervals along the contours in Fig.~\ref{fig1}. This is true independently of $\tau$. Notice the contrast here between correlation functions with the contour ordering $V W(t) V W(t)$ (which decay at large $t)$ and correlation functions with the ordering $V V W(t) W(t)$ (which do not).

The second piece of intuition requires us to introduce the thermofield double state in the Hilbert space of two copies of the quantum system, $|TFD\rangle = Z^{-1/2} \sum_n e^{-\beta E_n/2}|\bar n\rangle_L|n\rangle_R$. For any operator $O$, we define $O_L= O^T\otimes 1$, acting only on the $L$ Hilbert space, and $O_R = 1 \otimes O$ acting only on $R$. As an entangled state, $|TFD\rangle$ has a very nongeneric pattern of correlation between $L$ and $R$. In particular, simple operators are highly correlated, so that e.g. $\langle TFD|V_L V_R|TFD\rangle$ is large. 

The point of this preparation is that we can understand $F(t)$ as a similar two-sided correlation function in a perturbed version of $|TFD\rangle$. Specifically, $F(t) = \langle \Psi| V_L V_R|\Psi\rangle$ where
\be
|\Psi\rangle = Z^{-1/2}\sum_{mn}e^{-\beta(E_m+E_n)/4}W(t)_{n \bar m}|\bar m\rangle_L| n\rangle_R.
\ee
For small $t$, the simple $W$ operator will not significantly change the global pattern of correlation in the state, so $F$ remains large. However, as $t$ increases, the $W(t)$ perturbation becomes more and more complicated, and the delicate local correlations present in the thermofield double state will be destroyed \cite{SS}, causing $F$ to become small. This perspective makes the connection to the classical butterfly effect particularly clear.

Note that from the two-sided perspective, the ordering of operators in $F$ is quite natural. $F(t \pm  i \beta/4) $ has a simple interpretation as a correlator in the thermofield double state, with two operators acting on one side and two operators acting on the other. This correlation function is actually time ordered with respect to a two-sided time that increases forwards on both sides. However, in the rest of the paper we will reserve the term ``time ordered'' for configurations where the order of the operators in the trace coincides with the order expected when we view $t$ as a time variable. 
The distinction is important because $t$ runs forwards on the $R$ system and backwards on $L$.

Another important class of examples with a large hierarchy between scrambling and dissipation scales are the large $N$ gauge theories and related systems that can be studied using gauge/gravity duality.  Here the number of degrees of freedom is $N^2 = 1/\epsilon.$ For such systems, there has been recent progress in computing correlators such as $C(t)$ and $F(t)$ using holographic techniques in black hole backgrounds  \cite{SS,Shenker:2013yza,RSS,kitaev,Shenker:2014cwa}.\footnote{See also \cite{Roberts:2014ifa,Jackson:2014nla} for computations using related large $c$ sparse spectrum techniques in $d=2$ CFTs.}  The key element of these calculations is the connection of the long time behavior of (\ref{C}) and (\ref{F}) to a high energy scattering process near the bulk black hole horizon. The center of mass energy squared $s \sim \frac{ 1}{\beta^2}  \exp{\frac{2 \pi}{\beta} t}$ grows exponentially with time, as dictated by the local Rindler structure of the horizon \cite{'tHooft:1990fr,Kiem:1995iy}. The strength of this scattering becomes of order one when $G_N s \sim 1$ in AdS units.

For a large $N$ CFT holographically described by Einstein gravity, the methods of \cite{SS,Shenker:2013yza,RSS,kitaev,Shenker:2014cwa} give (for $t\gg \beta$)
\be\label{gravres}
F(t) = f_0 - \frac{f_1}{N^2} \exp{\frac{2 \pi}{\beta} t} + {\cal O}(N^{-4})
\ee
where $f_0,f_1$ are positive order one constants that depend on the specific operators $V,W$. The growing $N^{-2}$ term gives the first indication of the butterfly effect, that is, the beginning of a rapid decrease of $F(t)$ that takes place near the scrambling time $t_* = \frac{\beta}{2\pi}\log N^2$.  The dissipation time in such systems is determined by black hole quasinormal modes which give $t_d \sim \beta$ for low dimension operators.  So again there is a large hierarchy between scrambling and dissipation.

This result provides the reference point for the following conjecture.

\section{Conjecture}\label{conjsec}
We conjecture that chaos can develop no faster than the Einstein gravity result (\ref{gravres}) in thermal quantum systems with many degrees of freedom\footnote{In semiclassical billiards this would be the number of cells in phase space.}  and a large hierarchy between scrambling and dissipation.This conjecture is similar in spirit to the $\eta/S$ result of KSS \cite{Kovtun:2004de} that points to black holes in Einstein gravity as systems with very strong scattering.  It is a refinement of the fast scrambling conjecture of \cite{Sekino:2008he} which again singles out black holes.

In such systems, out of time order correlators such as $F$ in \nref{F} should display the following behavior. Well after the dissipation time  $t_d$, but well before the scrambling time $t_*$ they take an 
approximately constant factorized value $F(t)\approx F_d$, where
\be \la{factorization}
F_d \equiv  \tr[ y^2 V y^2 V] \tr[ y^2 W(t) y^2 W(t)] 
\ee
is the product of disconnected correlators. Due to time translation invariance, this is independent of $t$. For example, in a large $N$ system with $t$ independent of $N$, large $N$ factorization implies
\be
F(t) = \left(\tr[V y W(t) y^3]\right)^2 + \left(\tr[W(t) y V y^3]\right)^2 + \tr[y^2 V y^2V]\tr[y^2W y^2W] + {\cal O}(N^{-2})
\ee
The first two terms decay to zero for $t \gg t_d$.    In more general systems the role of large $N$ is played by the large number of degrees of freedom that cause commutators to be small for $t \ll t_*$.  

However, due to quantum mechanics and chaos, $F(t)$ cannot remain a constant forever. Scrambling causes a commutator to develop and $F(t)$ to decrease.  We conjecture that this rate of decrease is bounded (for times greater than a time $t_0$, which will be discussed at length in \S~\ref{arg}):
\be
\frac{d}{dt}(F_d - F(t)) \le \frac{2\pi}{\beta}(F_d - F(t))~.\label{cona}
\ee
  As Kitaev \cite{kitaev} has emphasized, building on \cite{larkin}, if the system is chaotic   we expect correlators like $F_d -F(t)$   
to  initially grow exponentially 
\be
 F_d - F(t) =  \epsilon  \exp{\lambda_L t} + \cdots  \label{lyapunov}
 \ee
 where $\lambda_L$ might depend on the operators $V,W$ as well as the particular quantum system.  We will follow Kitaev and refer to $\lambda_L$ as a Lyapunov exponent. (This  exponential behavior and the factor of $\epsilon$  in \nref{lyapunov} are related to the fast scrambling conjecture of \cite{Sekino:2008he}.)

Assuming this form we conjecture  the existence of a  universal bound 
\be \la{lybd}
\lambda_L \leq { 2 \pi \over \beta } = 2 \pi T~ .
\ee
In the following section we present evidence motivating this bound.  In \S~\ref{arg} we give a precise argument, based on plausible physical assumptions,  establishing it.

\section{Motivation for the conjecture}\label{evidence}
A number of lines of evidence led us to this conjecture. These involve  the study of large $N$ gauge theories, with and without gravity duals. 
\subsection*{Einstein gravity}
In the holographic calculations \cite{SS,Shenker:2013yza,RSS , kitaev,Shenker:2014cwa} that use Einstein gravity in the bulk, the result (\ref{gravres}) holds independent of $d$ and independent of the choice of $V$ and $W$.  This is because {\it (i)} gravitational scattering is of order $G_N  s$ (in AdS units) because the graviton is spin two, and $G_N \propto N^{-2}$, {\it (ii)} gravity couples universally,  and {\it (iii)} $s \sim \exp{\frac{2 \pi}{\beta} t}$ because of the kinematics  of Rindler horizons.  

\subsection*{Higher derivative corrections}
The result (\ref{gravres}) is unchanged if Einstein gravity is modified by higher derivative corrections  with a finite number of derivatives, like the Gauss-Bonnet term \cite{kitaev}. This is because such corrections do not change the spin of the graviton, so {\it (i)} remains true. The relation {\it (iii)} also remains correct as long as the thermal state is dual to a black hole with a smooth horizon. Notice that the situation here is different than for the $\eta/S$ calculation, where higher derivative couplings can move $\eta/S$  above and below the reference Einstein value of $1/4 \pi$ \cite{Kats:2007mq,Brigante:2007nu,Brigante:2008gz}.  This suggests that a sharp bound might exist for $\lambda_L$.

\subsection*{Weak coupling}
If the gauge theory is weakly coupled, with `t Hooft coupling $\lambda$ independent of $N$, the intuition described in \cite{Sekino:2008he} suggests that because the strength of gluon scattering  in the gauge theory is of order $\lambda$ at small $\lambda$,   the Lyapunov exponent  should be small, $\lambda_L \sim \lambda/\beta$, parametrically smaller than in the gravitational limit.\footnote{
For the particular case of Rindler AdS black holes (hyperbolic black holes at temperature $\beta = 2\pi$) it was indicated in \cite{Shenker:2014cwa} that $\lambda_L$ is the same as the Regge intercept $ j(t=0)-1$ in the gauge theory, which can be computed
using the BFKL analysis and is of order $\lambda$ at small $\lambda$.  See the discussion in \cite{Banks:2009bj}. This case is discussed in more detail in Appendix A.   It was also suggested in \cite{Shenker:2014cwa} that  a modification of the BFKL weak coupling calculation would allow the calculation of $\lambda_L$ at weak coupling in more general cases. }  We expect this to be true in any weakly coupled theory. 

\subsection*{Stringy corrections}
In a bulk weakly coupled string theory in a geometry with large radius of curvature, the first corrections to the Einstein gravity calculation of scrambling can be computed using the perturbative string theory techniques of \cite{Brower:2006ea}.   For planar or spherical horizons, Ref.~\cite{Shenker:2014cwa} showed that 
\be
\lambda_L = \frac{2 \pi}{\beta}\left[1 - {1 \over  2}\mu^2 l_s^2+...\right].\la{stringy}
\ee
where $\l_s$ is the string length, and $\mu^2$ is a constant that appears in the equation for a shock wave propagating along the horizon. This  equation involves the transverse dimensions and is of the form $(\nabla_{\perp}^2 - \mu^2) h =0$.  Here $h$ is the shock wave profile.\footnote{For the particular case of a
 planar $AdS_{d+1}$ black brane we have $\mu^2 = { d (d-1) \over 2 \ell_{AdS}^2 } $ \cite{Shenker:2014cwa}.} (If the horizon is hyperbolic, we replace $\mu^2$ by $\mu^2 + k_0^2$ in \nref{stringy}, where $k_0^2$ is the lowest eigenvalue of $-\nabla_{\perp}^2$.)  Einstein's equations imply that $\mu^2$ is positive if the transverse space warp factor grows as one moves away from the bifurcation surface in a spacelike direction \cite{Dray:1984ha,Sfetsos:1994xa}.\footnote{We are grateful to Mark Mezei for discussions on this point.} This will be the case if this horizon corresponds to a wormhole-like configuration.   This is the geometry appropriate to the dual of a thermal field theory.  

\subsection*{Scattering bound}
Because of the Rindler relation between bulk scattering energy  and time $s \sim \exp{\frac{2 \pi}{\beta} t}$ the bound (\ref{lybd}) is equivalent to the bulk statement that the eikonal phase $\delta$ is of order $G_N s^{p}$ and $p \le 1$.  (A spin $J$ field exchanged in the Mandelstam $t$-channel gives $p = J-1$).  The authors of \cite{Camanho:2014apa} argued that in scattering $p$ must be $\le 1$ because causality requires $e^{i\delta(s)} $ to be analytic in the upper half of the complex $s$ plane and unitarity requires  $|e^{i \delta(s)}| \le 1$ there.  This is  consistent with our conjectured bound and suggests that unitarity, analyticity and causality are the crucial assumptions necessary to prove the bound.  We work in Hamiltonian systems where unitarity and causality are manifest, and correlation functions are analytic. Because of the relation between $s$ and time $t$, a natural strategy is to formulate a bound on $F$ in the complex $t$ plane.

\section{Argument}\label{arg}
In this section we will provide a two-part argument for the bound. The first part consists of a simple mathematical result bounding the derivative of any function that satisfies certain assumptions. The second part consists of physical arguments that $F$ should satisfy closely related assumptions in the systems of interest. The resulting bound on the derivative implies \nref{cona}.

\subsection{A mathematical result} \label{math}
Suppose we have a function $f(t)$ with the following properties:

\begin{enumerate} 

\item
$f(t + i\tau )$ is analytic in the half strip $0<t$ and $ - {\beta \over 4} \leq \tau \leq { \beta \over 4 }$. (Here $t$ and $\tau$ are the real and imaginary parts of the complex number $t + i \tau$.)  We also assume that $f(t)$ is real for $\tau = 0$.

 \item 
$ |f(t+ i \tau)| \leq 1 $ in the entire half strip.

 \end{enumerate} 
Then we claim that 
\be \la{Prebd}
 { 1 \over 1 - f } \left| { d f \over d t} \right| \leq   { 2 \pi \over \beta }     + {\cal O}(e^{ - 4 \pi t/\beta}).  
 \ee
 
 Before presenting the proof, it is useful to consider the example $f(t) = 1 - \epsilon e^{\lambda_L t}$.   Here it is easy to see that the above properties imply the bound $\lambda_L \le { 2 \pi \over \beta }$.

To establish the claim in general, we first map the half strip to the unit circle in the complex plane using the transformation 
 \be
z = \frac{1 - \sinh \left[\frac{2\pi}{\beta}(t + i\tau) \right] }{1 + \sinh \left[ \frac{2\pi}{\beta}(t + i \tau) \right] }.\label{uniform}
\ee
 Then $f(z)$ is an analytic function from the unit disk into the unit disk, thanks to the second property. 
Such functions cannot increase  
 distances in the hyperbolic metric (the Schwarz-Pick theorem). The hyperbolic metric is $ds^2 = 4dzd\bar{z}/(1 - |z|^2)^2$, so we must have
\be
\frac{|df|}{1 - |f(z)|^2} \le \frac{|dz|}{1-|z|^2}.
\ee
We apply this inequality for $\tau = 0$ where $f$ is real, finding 
\be \label{bdcoth}
{ 1 \over 1 - f } \left| { d f \over d t} \right| \leq { 2 \pi \over \beta} \, \coth\left( { 2 \pi t \over \beta} \right) \,  { ( 1 + f ) \over 2 }   \leq { 2 \pi \over \beta}  + {\cal O}(e^{ - 4 \pi t/\beta } ) 
\ee
which is the claim \nref{Prebd}.

\subsection{Deriving the bound} 
If we could show that $F(t)/F_d$ satisfies properties one and two, above, then \nref{Prebd} would imply the conjecture \nref{cona}. Recall that $F_d$ is the disconnected correlator
\be
F_d\equiv \tr[ y^2 V y^2 V] \tr[ y^2 W y^2 W].
\ee

The first property is easy to establish. The meaning of $F$ for complex times is most simply understood from Fig.~\ref{fig1}, but we can also write it out explicitly as
\be\label{Ftau}
F(t + i\tau) = \frac{1}{Z}\tr[ e^{-(\beta/4- \tau)H}Ve^{-(\beta/4 + \tau)H}W(t)e^{-(\beta/4- \tau)H}Ve^{-(\beta/4 + \tau)H}W(t)].
\ee
For finite $N$ and finite volume the RHS defines an analytic function in the strip $|\tau| \le  \beta/4$, even in quantum field theory. We also see that when $\tau =0$ $F(t)$ is real. (Recall that $W$ and $V$ are Hermitian operators.) Therefore the first property holds in general. 

The second property is more subtle. In fact, we will only show that $|F(t+i\tau)|\le F_d + \varepsilon$, for an appropriate $\varepsilon$, and for times $t$ greater than a reference time $t_0$. This will allow us to apply the result from the previous section to the function
\be
f(t) = { F( t+ t_0) \over F_d + \varepsilon }.\label{actualf}
\ee
Provided that $\varepsilon$ is small, this will give us the bound \nref{cona} up to small errors, for times greater than $t_0$. We will derive conditions on $\varepsilon$ and $t_0$ in the process of arguing that $f$ satisfies property two.

Our strategy will be to show that $|f(t+i\tau)|\le 1$ on the three boundaries of the half strip $0<t$ and $-\beta/4 < \tau < \beta/4$, and that $f$ is bounded by some constant everywhere in the interior. Then the Phragm\'{e}n-Lindel\"{o}f principle (the analog of the maximum principle for non-compact regions) implies that the function actually obeys $|f(t+ i \tau) |\leq 1$ everywhere in the interior, establishing the second property. 

First, we consider the edges of the half strip $|\tau| = \beta/4$. Notice that
\be
F(t - i\beta/4) = \tr[y^2 V W(t) y^2 V W(t)].
\ee
The RHS can be viewed as an inner product of ``vectors'' $[y V W(t) y]_{ij}$ and $[y W(t) V y]_{ij}$ ($W$ and $V$ are assumed Hermitian). The Cauchy-Schwarz inequality then gives\footnote{Note that at leading $N^{-2}$ order, the Einstein gravity result (\ref{gravres}) saturates this bound.}
\be
|F(t-i\beta/4)| \le \tr[y^2 W(t) V y^2 V W(t)].   \label{CS}
\ee
In a chaotic system with many degrees of freedom, and for times large compared to the dissipation timescale, we expect that the RHS factorizes and is given by $F_d$. This is the main physical input to the argument. To make the possible error explicit, we define $\varepsilon$ by the condition that for all $t \ge t_0$, we will have
\be \label{errorest}
\tr[y^2 W(t) V y^2 V W(t)] \le \tr[ y^2 W(t) y^2 W(t)] \tr[y^2 V y^2 V] + \varepsilon ~.
\ee
In general the size of $\varepsilon$ will depend on $t_0$. In systems where we can take $\varepsilon$ small while keeping $t_0 \ll t_*$, we will get a good approximation to the bound \nref{cona} once $F_d - F(t)$ exceeds $\varepsilon$. We will analyze $\varepsilon$ and $t_0$ in some example systems in the following sections. For the present purposes, the important point is that with the definition of $\varepsilon$ in \nref{errorest}, the Cauchy-Schwarz inequality ensures that $|f|\le 1$ on the edges $|\tau| = \beta/4$. 

Next, consider the third boundary at $t = 0$. This corresponds to $F(t_0 + i\tau)$ with $-\beta/4 \le \tau \le \beta/4$. Here the possible error in factorization has two sources. One is the failure of the time-ordered correlation function to factorize, which is order $\varepsilon$. The other is due to the fact that $F$ is not time-ordered; $F$ will begin to move from its factorized value due to the onset of scrambling. In general, we expect this to cause $F$ to decrease, but it is not necessary to assume this. As long as we choose $t_0$ early enough that the effect of scrambling is smaller than the $\varepsilon$ defined by condition \nref{errorest}, the second error will be smaller than the first, so $|f|\le 1$ on the third boundary as well.

To complete the argument for property two via the Phragm\'{e}n-Lindel\"{o}f principle, we need to establish that $f$ is bounded in the interior by some constant, $|f(z) |\leq C$, where $C$ might be bigger than one. Again, we apply the Cauchy-Schwarz inequality, viewing $F$ as the product of two vectors. Choosing the vectors appropriately we find (for positive $\tau$) 
\be 
\begin{split}
|F(t + i \tau ) |\le &   
 \tr[  y^{ 1 + \eta } V y^{ 1 - \eta} W y^{1 + \eta} W y^{1 - \eta } V ]  \\
 \sim & \tr[ y^{ 1 +\eta} V y^{ 3 - \eta} V] \tr[y^{1+ \eta} W y^{ 3 -\eta } W] \\ \leq & \tr[ y V y^2 V] \tr[ y W y^2 W ] \la{intbound}
 \end{split}
\ee
with $\eta = { 4 \tau \over \beta } $. In the second line we have again invoked factorization at late times for time ordered correlators. (All the $W$ are evaluated at time $t$ and $V$ at time zero.)  The third line uses Hermiticity of $V, W$ and the contracting property of $y$.   What appears on the RHS is not the same as $F_d$, since we have fewer powers of $y$ compared to \nref{errorest}, but it is finite, so we have established property two.
 
Finally, let us address a slight imprecision in our discussion.  In the above we assumed
that the largest times we would talk about are of order the scrambling time, which are logarithmic in the small parameter. On the other hand, after very large times we can have Poincare recurrences, and we expect factorization to fail. To avoid this we can cut off the half strip by adding an additional boundary at a time much larger than
the scrambling time but much smaller than the recurrence time. At this additional boundary we need to have $|F|\le F_d+\varepsilon$. In a chaotic system, we expect $F$ to be very small for almost all times, so it should be easy to find a suitable time for the cutoff.\footnote{Assuming incommensurate energies, one can show that the long time average of $F$ is exponentially small in the entropy of the system.} The conformal transformation from this finite strip to the disk will be more complicated, but it will coincide with the one we used in the region of interest for our arguments.

We conclude that $f$ in \nref{actualf} satisfies properties one and two from the previous section. The mathematical result \nref{Prebd} then implies that for $t$ greater than $t_0$ plus a few thermal times, we have
\be\la{result}
\frac{d}{dt}(F_d - F(t))\le \frac{2\pi}{\beta}(F_d-F(t) + \varepsilon).
\ee
Here, to recap, $\varepsilon$ is the maximum error in the time ordered factorization \nref{errorest} for times $t \ge t_0$. For different systems, we might make different choices of $\varepsilon$ and $t_0$, in order to get the best bound. Examples will be discussed below. The essential point is that for a wide class of chaotic systems where $V,W$ are small perturbations, we expect the scrambling time $t_*$ to be large, and we expect factorization to hold up to small errors after a time $t_0$ with $t_0 \ll t_*$. For such systems, the result \nref{result} implies the bound \nref{cona} for the growth of $F_d - F(t)$ once this quantity exceeds the small error.

\subsection{Examples}
\subsubsection{Large $N$ systems}
In large $N$ systems we can take $V$ and $W$ to be single trace operators and exploit large $N$ factorization.  The error $\varepsilon$ in the estimate discussed in \nref{errorest} is then given by
\be
\varepsilon = \left(\tr[V y W(t) y^3]\right)^2 + \left(\tr[W(t) y V y^3]\right)^2 + {\cal O}(N^{-2}) ~.
\ee
For general $V$ and $W$ these off diagonal expectation values are nonzero but decay because of dissipation, leading to an estimate $\varepsilon \sim N^{-2} + e^{-t_0/t_d}$. We must now choose $t_0$. In order to get the best bound, we set $\varepsilon$ equal to the growing effect of scrambling on $F(t)$ at time $t_0$. As an example, suppose that $F_d - F(t)$ is proportional to $\epsilon\, e^{\lambda_L t}$. Then the optimal $t_0$ is given by $t_*/(1+ \frac{1}{\lambda_L t_d})$, and we have $\varepsilon \sim \epsilon^{\lambda_L t_d/(1 + \lambda_L t_d)}$. Once $F_d - F(t)$ exceeds this value (near the time $t_0$), \nref{result} implies the bound \nref{cona}.

We can get a bound for a wider range of times if the system has a global symmetry like parity and we choose  $V$ and $W$ to  transform differently under it. Then the first two terms above vanish and $\varepsilon \sim N^{-2}$.  This means we can take $t_0 = 0$ and still make chaotic effects dominate over the error.  Because chaos is almost by definition generic even special operators will couple to the basic chaotic dynamics of the system so we can apply the bound to the very early development of this chaos.  In particular, by integrating \nref{bdcoth} from early time we find
\be
F_d - F(t) \le { c \over N^2}   \exp \left( { { 2 \pi \over \beta} t }  \right)
\ee
where $c$ is an order one $N$ independent constant.
  
\subsubsection{Extended local systems} 
For lattice systems, or for thermal quantum field theories, the large number of degrees of freedom comes from the fact that we have an extended system. We can take $V$ to be an operator at the origin and $W$ to be an operator at a site at large distance $L$. For such systems, we get an interesting bound by setting $t_0 = 0$. Then $\varepsilon$ is equal to the maximum over $t$  of
\be
\tr[y^2 W(t) V y^2 V W(t)]-\tr[ y^2 W(t) y^2 W(t)] \tr[y^2 V y^2 V].\la{difference}
\ee
At $t = 0$, we expect the above to be $\sim e^{-c_1 L}$ in general, because of the short range correlations in the thermal state. 

In special systems (such as those discussed below), this factorization might break down for times $t\propto L$, due to the possibility of signalling between $W$ and $V$. However, for generic chaotic systems at finite temperature, we expect that signals should be exponentially suppressed in distance, so that the difference in \nref{difference} is $\le e^{-c_2 L}$ for all time. We can then take $\varepsilon = e^{-c_2 L}$. As before, the bound \nref{result} implies \nref{cona} once $F_d - F(t)$ exceeds this small value. Note that this may take a long time if $L$ is large.

\subsubsection{Cases where there is no bound} \label{casesnbd}
There are local systems for which factorization $\langle V W(t) W(t) V\rangle\approx \langle VV\rangle \langle WW\rangle$ does not hold for an appropriate range of times, even for widely separated $V,W$. For example, consider a massless free field $\phi$ in two dimensions and take the operators to be $V(0) = \partial_- \phi(0)$ and $W(t) = \partial_- \phi(L - t)$. Even if $L$ is large, the contraction between $V$ and $W$ becomes important for $t \approx L$. This is the same time at which the commutator becomes nonzero, so we cannot bound its growth.

Indeed, in this system the commutator is $[V, W(t) ] \propto \delta'(L - t )$, which rises very fast, independently of the temperature.\footnote{We can smear the operators a bit, but we retain the same conclusion: the growth is determined by the parameters of the smearing function rather than the temperature.}

There is a related issue in any two dimensional conformal field theory. Such theories contain a stress tensor operator
$T_{--}(x^-)$ which has singularities along the light-cone. Taking $V = T_{--}$ and $W$ some other local operator we find that factorization fails near the light cone. However, at large $c$, this is suppressed by $1/c$ and (after smearing) can be absorbed within the small $\varepsilon$ that we are tolerating.

%The motivation for the assumptions was that we had a chaotic system, however, in this case we have an exactly conserved quantity, and we do not expect the assumptions to hold. This is not an issue when the CFT has large central charge because in that case the coupling to the stress tensor operator is suppressed by $1/c$ and such effects are within the small corrections that we were tolerating.

In fact this is a problem specific to two dimensional systems where the light cones are one dimensional so signals cannot spread around them.
In higher dimensions this is not an issue and the commutators are suppressed at large spatial separation. This is easy to see in free theories, and for general conformal field theories, as explained in Appendix A.

In addition to the factorization assumption, we have assumed that there is a large hierarchy between the dissipation time and the scrambling time. We have justified this on the grounds that we have many degrees of freedom and that the Hamiltonian is built from finite products of simple operators. Alternatively, one could consider a Hamiltonian given by a random Hermitian matrix. For such a system, we expect no such hierarchy, so our conjecture does not apply.
%In conclusion, the physical assumptions we made are not always true. But we expect them to be true in generic chaotic systems.

 \subsubsection{Rindler space and the scattering bound} 
Field theories on Rindler space are simple examples of thermal systems.  
In this case the Minkowski vacuum is the 
thermofield double state. For the case of conformal field theories in $d>2$, one can prove that the bound \nref{result} holds with small $\varepsilon$ for Rindler correlators of well separated operators. This follows from the fact that the correlators are
related to Minkowski vacuum four point functions which can be approximated using the operator product expansion. 
For theories with gravity duals this implies the scattering bound \cite{Camanho:2014apa} mentioned in \S~\ref{evidence}. 
We discuss this point more extensively in appendix A. This appendix also serves as a worked
out example of  the considerations in this paper. 
 
  \subsubsection{Semiclassical billiards} 
At first sight one might think that a classical system could violate the bound since classical
Lyapunov exponents can take any value. However, restoring dimensionful factors, our conjecture is 
\be
\lambda_L \le { 2\pi k_B T \over  \hbar},
\ee
so there is no contradiction in the strict classical limit $\hbar \rightarrow 0$.

It is interesting to consider a semiclassical chaotic system with a small $\hbar$ at finite temperature. For such systems, we can take $\varepsilon \sim e^{-t_0/t_d}$ as with the large $N$ case. The analysis is as before. One can also give a direct (although heuristic) argument for a bound, following reasoning in \cite{Kovtun:2004de}.  Consider a semiclassical chaotic system such as interacting quasiparticles or stadium billiards. A naive definition of the Lyapunov exponent is the inverse of the timescale $\tau_{\rm nl}$ over which the evolution of a particle becomes 
  nonlinear. For example, $\tau_{\rm nl}$ would be proportional to the mean free time for a system of interacting quasiparticles, 
  or the time to cross the stadium for a billiards problem. To violate the bound, we would need 
$ \tau_{\rm nl } \, k_B T \lesssim \hbar$.
  Since $k_B T$ is the typical energy, 
   we would need a violation of the energy-time uncertainty principle, indicating that the semiclassical description is invalid.

   \section{Concluding remarks}
We have given a strong argument for a bound on the rate at which chaos can develop in general thermal quantum systems with a large number of degrees of freedom. The large number of degrees of freedom suppresses the initial size of the commutator causing strong chaos--scrambling--to develop parametrically later  than dissipation.  We diagnosed chaos using an out of time order correlator $F(t)$ related to a commutator. Characterizing this growth in terms of a Lyapunov exponent, we claim that it is bounded by 
\be
 \lambda_L \leq { 2 \pi k_B T \over \hbar } 
\ee
where $T$ is the temperature of the system. 

Our direct argument for this bound relied on analyticity, as well as the physical input that certain time-ordered correlation functions should approximately factorize. We gave arguments justifying this factorization for different classes of physical systems with many degrees of freedom. In the general case, these arguments also relied either on large timelike or spacelike separation between operators.

It is tempting to speculate \cite{kitaev} that a large $N$ system which saturates this bound will necessarily have an Einstein gravity dual, at least in the near horizon region. This is in the spirit of the speculation in \cite{Heemskerk:2009pn} that a system with no light higher spin single trace states should have a gravity dual.

\section*{Acknowledgements}

 We thank A. Kitaev, M. Mezei, and A. Wall for helpful discussions. J.M. is supported in part by U.S. Department of Energy grant
de-sc0009988. S.S. is supported in part  by  NSF grant  PHY-1316699 and by a grant from  the John Templeton Foundation.  D. S.  is supported in part by NSF grant  PHY-1314311/Dirac.

\appendix 

\section{Rindler space and the scattering bound} 
 
\begin{figure}
\begin{center}
\includegraphics[scale = 0.5]{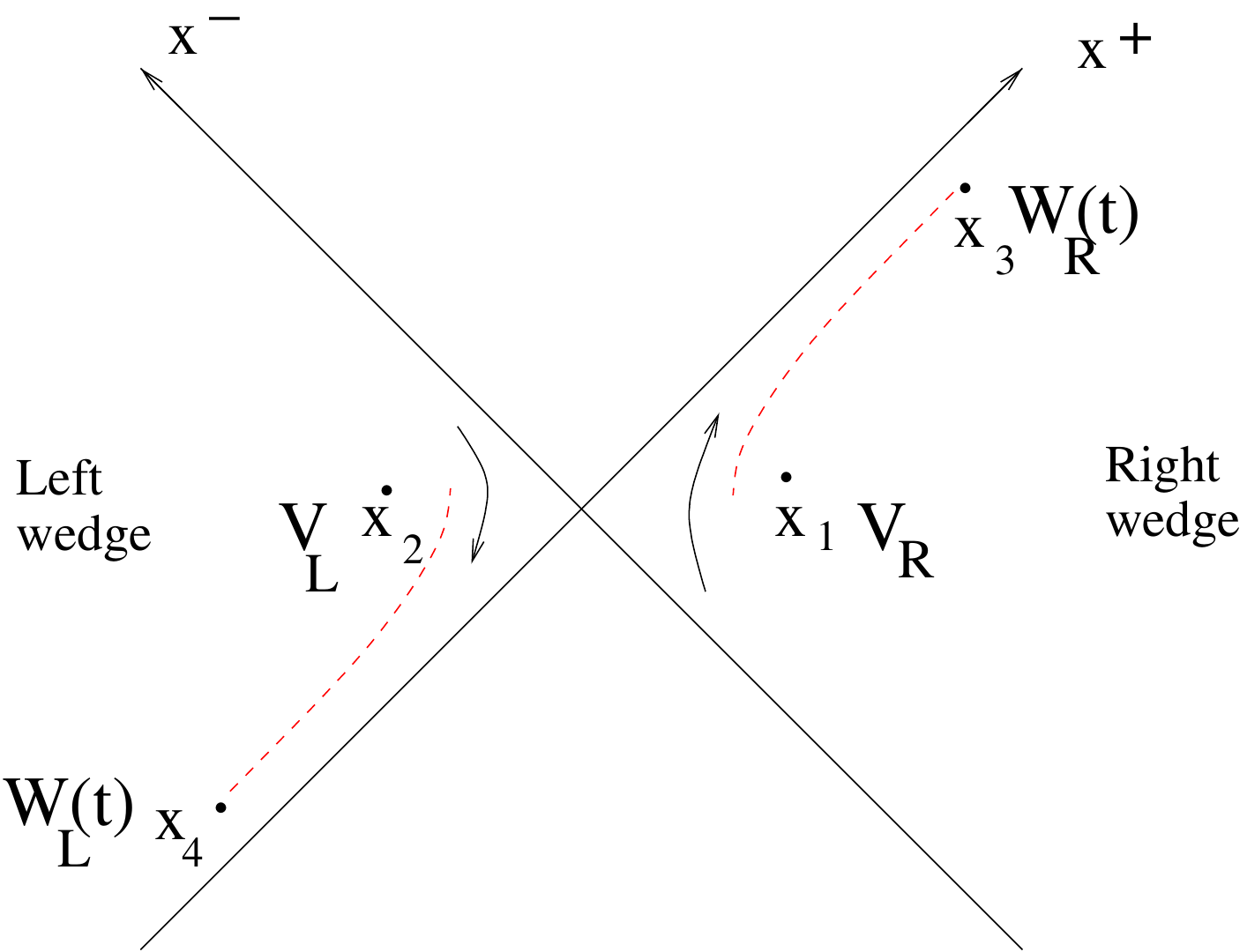}
\caption{ Consider a two dimensional Minkowski space, with its right and left Rindler wedges. We insert two operators in the right Rindler wedge and two on the left. 
Then we act on $W_L$ and $W_R$ by the Rindler time translation generator, which is a boost around the origin. This translates $W_R$ upwards in time and 
$W_L$ downwards in time, as shown by the arrows.  }\label{PointArrangement}
\end{center}
\end{figure}

The Rindler construction gives simple examples of thermal systems. We  consider a CFT${}_d$ on Minkowski space and choose Rindler coordinates 
$ds^2 = -\rho^2 dt ^2 + d\rho^2 + d\vec x^{\, 2}_{d-2} $. 
%This metric is Weyl equivalent to 
% $ds^2  = -dt ^2 + d H_{d-1}^2 $. 
The Minkowski vacuum corresponds to a thermal state on Rindler space. These coordinates cover the right 
 Rindler wedge. There is an identical set of coordinates which cover the left Rindler wedge, see figure \ref{PointArrangement}.
 The Minkowski vacuum can be viewed as the thermofield double, entangling these two systems.
 We can now apply our general discussion to the particular case of a Rindler wedge. 
 In this context the function $F(t\pm i \beta/4)$
 corresponds to an ordinary Minkowski space four point function. More precisely, imagine that we choose all four points inside a two dimensional $R^{1,1}$ subspace
 of the full $R^{1,d-1}$ space. Let us insert the four operators as shown in figure \ref{PointArrangement}, with the points 
 \be \la{FourPoints}
 % x_1^\pm = \pm e^{-\sigma/2} ~,~~~~x_2^\pm = \mp e^{ -\sigma/2} ~,~~~~~~x_3^\pm = \pm e^{ \sigma/2   } e^{\pm t}~,~~~~~~ x_4^\pm  = \mp e^{ \sigma/2 } e^{\pm t } 
  x_1^\pm = \pm 1 ~,~~~~x_2^\pm = \mp 1 ~,~~~~~~x_3^\pm = \pm e^{ \sigma \pm t}~,~~~~~~ x_4^\pm  = \mp e^{ \sigma \pm t } 
   \ee
The cross ratios then become 
\be 
  z_\pm = { x_{12}^\pm x_{34}^\pm \over x_{14}^\pm x_{32}^\pm }   = {  1 \over \cosh^2({  t \pm \sigma \over 2 }) } 
 \ee
 Here we have used the label $t$,  as in the rest of this paper, to denote the flow by the Killing vector generating Rindler time translations. Note that this flows 
 backwards in time on the left Rindler wedge, see figure \ref{PointArrangement}.

  All four point functions of conformal primaries $V,W$ can be computed by analytically continuing the flat space euclidean correlator, 
   with suitable $i \epsilon$ prescriptions \cite{Cornalba:2006xk, Roberts:2014ifa}.  
  The $i\epsilon$ prescription that gives rise to the $F$ correlator is the one that is natural from the point of view of Minkowski space. 
  More precisely, the correlator  $F(t + i \beta/4)$ corresponds to a correlator in Minkowski space with the standard time ordering\footnote{ 
  Recall that the  $i\epsilon$ prescription for any ordered Minkowski correlator 
$\langle 0| O(x_n) \cdots O(x_2) O(x_1) | \rangle$ is that we add $x_i^0 \to x_i^0 - i \epsilon_i$
with $\epsilon_{i} \leq \epsilon_{i+1}$. Note, however, that the shift in Minkowski time  to $-i \epsilon$ in the left wedge translates into a shift 
into the $+ i \epsilon $ direction in the $t$ coordinate due to opposite flow of time there.   }, 
 see figure \ref{PointArrangement},
      \be \la{Fco}
 F(t + i \beta/4) =\tr[ W(t) V y^2 W(t) Vy^2]   =   \langle 0|    W_R(t - i \epsilon )     V_R    V_L    W_L(t- i\epsilon) | 0 \rangle 
\ee
where we have not bothered to introduce $i\epsilon$'s for operators that stay spacelike separated as we change $t$. 
 On the other hand, a correlator that   naturally factorizes at large times is given by 
 \be \la{Tco}
 \tr[ V W(t)  y^2 W(t) Vy^2]  = \langle 0|   W_L( t +   i \epsilon)  W_R(t -  i \epsilon)  V_R    V_L     | 0 \rangle 
 \ee
 For $t =0$ and  $\sigma \ll 0$ these correlators are equal. They are in   the  Euclidean OPE region in the $VV$ channel (or 12 channel).  
 We will now keep $\sigma $ fixed and increase $t$. 
 Increasing  $t$ we pass through a point where two of the operators are null separated at $t + \sigma =0$. At this point 
 $z_+=1$. This is a singular point for the four point function. By suitably smearing the operators we can remove the singularity. 
 Notice that, for $\sigma \ll 0$,  the other cross ratio, $z_-$, remains small throughout the discussion. Therefore, using the OPE in the $VV$ channel, 
  we can expand the 
 correlators  in a series  of the form
   \be  \label{opevv}
   \sum z_+^{\Delta + S} z_-^{\Delta - S} c_{\Delta , S}  
   \ee  where $\Delta$ and $S$ are the dimension and spin of the intermediate operators. 
 Since $z_-$ is small, after smearing in $z_+$, we can apply a uniform bound for this quantity  when $d>2$,  since unitarity implies that\footnote{The exceptions in two dimensions pointed out in \S~\ref{casesnbd} follow from the existence of operators with $\Delta = S$ there, like the stress tensor.}   $\Delta -S \geq { d -2 \over 2} $.
  This holds on the first sheet of the $z_+$ plane. The $i\epsilon$ prescription
 in \nref{Tco} implies that  $z_+$ remains on the first sheet as we change $t$. But   for \nref{Fco} we circle around the branch cut at $z_+=1$, 
  which changes the behavior
 when we return to $z_+\to 0$.   In conclusion, we find that by taking $V$ and $W$ far away in space we ensure that \nref{Tco} factorizes as indicated in  (\ref{errorest})
 for all times.  
 Therefore the bound  (\ref{result})
  is a theorem  in this situation. 
  
  The dissipation time $t_d$ is just the inverse of the smallest $\Delta$ in \nref{opevv}. The manifest lack of recurrences here can be interpreted thermally as due to the infinite entropy of the thermal system on $H_{d-1}$.
  As we remarked above, here  the Lyapunov exponent is the same as the BFKL intercept $ \lambda_L = j (t=0) -1$   \cite{Banks:2009bj,Shenker:2014cwa}. 
  The high energy nature of the process for large $t$ is apparent from figure \ref{PointArrangement}.
  
 We now consider large $N$ CFTs which have an Einstein gravity dual.  
 We can extend the Rindler coordinates through the bulk and we can view the resulting space as a zero mass hyperbolic black hole, or a two sided hyperbolic black hole.    
 The bulk scattering that is dual to chaos here is just high energy gravitational scattering in vacuum AdS space.  More precisely $F$ is computed by folding bulk to boundary propagators against the bulk gravitational scattering amplitude \cite{kitaev, Shenker:2014cwa}.  When the scattering is weak\footnote{
 The parts of the propagators that correspond to strong scattering make a small contribution to $F$, which is dominated by $G_N s \sim 1$  at large  boundary time $t$ \cite{Shenker:2014cwa}.  So this argument for the bound only applies in the region where $G_N s$ is small (but order one).  } the propagator 
 variation is a small effect and the rate of decrease of $F$ directly diagnoses the size of the eikonal phase $\delta(s)$.  
  The bound (\ref{result}) shows that this phase cannot increase faster than $s$.

 This is an alternate derivation of the scattering bound in \cite{Camanho:2014apa} that helped motivate this work. 
  More precisely,  we get the bound  $| 1 + i \delta(s) |\leq 1 + {\cal O}(\delta^2) $  in the upper half $s$ plane, when $\delta(s)$ is small but of order one.
     This bound also  implies  the positivity of the Shapiro time delay. This is a nontrivial constraint for classical Gauss-Bonnet theories, it rules them out 
 as classical theories \cite{Camanho:2014apa}.    The exchange of a spin $J$ field in the Mandelstam $t$ channel gives $\delta(s) \sim s^{J-1}$.  Then 
 the bound  (\ref{result}) rules out any weakly coupled large radius bulk theory with a finite number of light particles with spin greater than two.

\end{document}